\documentclass[12pt]{iopart}

\usepackage{iopams}  

\usepackage{amsmath,amsfonts}
\usepackage{graphicx}

\newcommand{\half}{\textstyle{\frac{1}{2}}}
\newcommand{\dd}{\mathrm{d}}

\newcommand{\bbLL}{\mathbb{L}}
\DeclareMathOperator{\sech}{sech}

\begin{document}

\title[Billiards \&\ Ballyards]{Integrable elliptic billiards and ballyards}

\author{Peter Lynch}
\address{School of Mathematics and Statistics, University College, Dublin}
\ead{\mailto{Peter.Lynch@ucd.ie}}
\vspace{10pt}
\begin{indented}
\item[]July 2019 \\
\submitto{\EJP} 
\end{indented}

%%  \emph{Orcid ID:} Peter Lynch: http://orcid.org/0000-0002-8255-3404

%%%%%%%%%%%%%%%%%%%%%%%%%%%%%%%%%%%%%%%%%%
%%%%%%%%%%%%%%%%%%%%%%%%%%%%%%%%%%%%%%%%%%

%%%%%%%%%%%%%%%%%%%%%%%%%%%%%%%%%%%%%%%%%%%%%%%%%%%

\begin{abstract}
The billiard problem concerns a point particle moving freely in a
region of the horizontal plane bounded by a closed curve $\Gamma$,
and reflected at each impact with $\Gamma$.  The region is called
a `billiard', and the reflections are specular: the angle of
reflection equals the angle of incidence.  We review the dynamics
in the case of an elliptical billiard.  In addition to conservation
of energy, the quantity $L_1 L_2$ is an integral of the motion,
where $L_1$ and $L_2$ are the angular momenta about the two foci.
We can regularize the billiard problem by approximating the
flat-bedded, hard-edged surface by a smooth function.  We then
obtain solutions that are everywhere continuous and differentiable.
We call such a regularized potential a `ballyard'.  A class of
ballyard potentials will be defined that yield systems that are
completely integrable.  We find a new integral of the motion that
corresponds, in the billiards limit $N\to\infty$, to $L_1 L_2$.
Just as for the billiard problem, there is a separation of the
orbits into boxes and loops.  The discriminant that determines the
character of the solution is the sign of $L_1 L_2$ on the major
axis.


\end{abstract}

%%%%%%%%%%%%%%%%%%%%%%%%%%%%%%%%%%%%%%%%%%%%%%%%%%%%%%%%

\section{Introduction}
\label{sec:intro}

In his \emph{Lectures on Theoretical Physics}, Arnold Sommerfeld \cite{Som64}
wrote
``The beautiful game of billiards opens up a rich field for applications
of the dynamics of rigid bodies. One of the illustrious names in
the history of mechanics, that of Coriolis, is connected with it.''
Sommerfeld was referring to a book by Gaspard-Gustave de Coriolis,
\emph{Th\'{e}orie math\'{e}matique des effets du jeu de billiard},
published in Paris in 1835 \cite{Cor35}.

Billiards has been used to examine questions of ergodic theory
\cite{Bir42}.  In ergodic systems, all configurations and momenta
compatible with the total energy are eventually explored.  Such questions
lie at the foundation of statistical mechanics.
We know that the dynamics on an elliptic billiard is integrable: its
caustics are confocal ellipses and hyperbolas.  George Birkhoff conjectured
that if the neighbourhood of a strictly convex smooth boundary curve
is foliated by caustics, then the curve must be an ellipse.  So far,
this conjecture --- that ellipses are characterized by their integrability ---
remains an open problem.

%%  One may also investigate higher-dimensional billiards, for example, the
%%  motion of a particle confined within a spherical ball.  In this case,
%%  each trajectory lies within a plane.

The simplest billiard is circular. The dynamics are easily described:
every trajectory makes a constant angle with the boundary and is tangent
to a concentric circle within it.  
Berger \cite[p.~713]{Ber05} observed that ``Circular billiards is indeed a banal subject,
but nontheless we visualize it rapidly.'' Every trajectory is either a
polygon (perhaps star-shaped) or is everywhere dense in an annular region.
Moreover, the elliptical billiard problem is completely resolved, thanks to
Poncelet's theorem and the known geometry of confocal conics.
We have periodic trajectories, or ones that are dense in regions of two distinct topological types.
More generally, Tabachnikov \cite{Tab05}
discusses the theory of convex smooth billiards --- the elliptic
case.  Berry \cite{Ber81} considered various deformations of the circular
billiard, and showed how they support a variety of different orbits.

%%  Stadia give ergodic motion, exploring the entire energetically accessible
%%  domain.  Elliptical billiards have motions entirely confined to invariant
%%  curves whose topology is organised by two isolated closed orbits,
%%  and a family of ovals gives motion
%%  in which phase space is intricately divided into chaotic
%%  areas and areas covered with invariant curves.

In \S\ref{sec:billiards} we review the well-known theory of billiards on an elliptical
domain.  In this case, the velocity has a jump discontinuity at
each impact.  In \S\ref{sec:ballyards}, we introduce 
\emph{ballyard} domains, regularizing the problem by approximating the 
flat-bedded, hard-edged surface by a smooth function.
This ensures that the solutions are everywhere smooth. 
We define a countably infinite class of ballyard potentials depending
on a parameter $N$, each of which is completely integrable.
For this class, we find a new
integral of the motion, $\bbLL$, that corresponds, in the billiards limit
$N\to\infty$, to $L_1 L_2$.  It follows from this integral and the nature
of the potential surface that the orbits split
into boxes and loops.  The discriminant that determines the character
of the solution is the sign of $L_1 L_2$ on the major axis.

%%%%%%%%%%%%%%%%%%%%%%%%%%%%%%%%%%%%%%%%%%%%%%%%%%%%%%%%%%%%%%%%%

\section{Elliptical Billiards}
\label{sec:billiards}

We idealize the game of billiards, assuming the ball is a point mass
moving at constant velocity between elastic impacts with the boundary,
or cushion, of the billiard table. The energy is taken to be constant.
The path traced out by the moving ball
may form a closed periodic loop or, more generically, it may cover
the table or part of it densely, never returning to the starting conditions.

The billiard table may be of any shape, but we generally assume that
it is either rectangular, like a normal table, or elliptical. In this
section, we examine the orbits for an elliptical table.
We assume for simplicity that the centre of the ellipse is at the origin,
the major axis coincides with the $x$-axis and the semi-axes are
$a$ and $b$. Thus, the boundary is described by the equation
$$
\frac{x^2}{a^2} + \frac{y^2}{b^2} = 1
$$
It is also useful to recall the parametric form
$$
x = a\cos\theta \,, \qquad y = b\sin\theta
$$
where $\theta$ is called the eccentric anomaly (note that $\theta$ is not
identical to the polar angle, which we denote by $\vartheta$).
The foci are at $(f,0)$ and $(-f,0)$ where $f^2 = a^2 - b^2$.
Defining the eccentricity $e$ by $e^2 = 1-(b/a)^2$, it follows that the foci are
at $(ea,0)$ and $(-ea,0)$.

\subsection*{Initial Conditions.}

%  The orbit is determined by the initial position and angle of motion.
Initial conditions for motion in a plane normally require four numbers, two for the position and two
for the momentum. Since we can assume without loss of generality that the ball moves at unit
speed, only the direction of the initial momentum is
required. Likewise, if the motion starts from the boundary, a single value determines the position.
We suppose a trajectory starts at a boundary point with eccentric anomaly $\theta_0$
and moves at an angle $\psi_0$ to the $x$-axis. The pair of values $\{\theta_0,\psi_0\}$
determines the entire motion.

Each bounce is a specular reflection: the angle between the normal to the
cushion and the incoming trajectory is equal to the corresponding angle for the
outgoing trajectory. The tangential component of velocity is unchanged
after impact, while the normal component reverses sign. Thus, the speed
remains constant.

The first linear segment of the trajectory is tangent to another conic confocal 
with the boundary. Because of the specular reflection at the boundary, all subsequent
segments are also tangent to this conic \cite{Ber05}.
The conic is called the caustic of the orbit, and may be an ellipse or a hyperbola.

\subsection*{Generic Motion: Box Orbits \&\ Loop Orbits.}

Since the speed is a constant, taken to be unity, the orbit is
determined by the initial postion and angle of motion. Alternatively,
we can simply give the first two points of impact with the cushion.
There are two generic types of orbit. The first arises when the ball crosses
the major axis between the foci, the second when it crosses outside.

\begin{figure}[h]
\begin{center}
\includegraphics[scale=0.667]{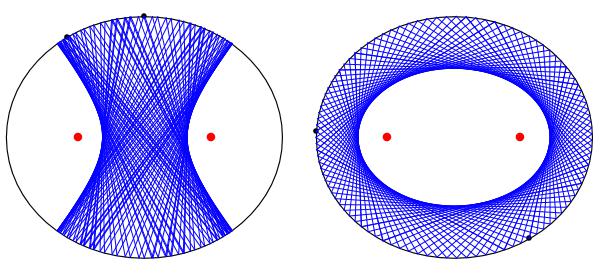}
\caption{Generic orbits. Left: Box orbit, crossing the major axis between the foci.
Right: Loop orbit, crossing the major axis outside the foci.}
\label{fig:genericorbits}
\end{center}
\end{figure}

In the first case, once the trajectory brings the ball across the
interval $-f < x < +f$ on the $x$-axis, it is constrained by the geometry of
the ellipse to remain within a region bounded by a hyperbola having the
same foci as the original ellipse. Each segment of the orbit between
impacts with the ellipse is tangent to this hyperbola
(see figure~\ref{fig:genericorbits}, left panel).
Generically, the orbit is dense in this region, coming arbitrarily close
to any given point within it.

If the ball crosses the major axis outside the foci, that is for $x\in(-a,-f)$
or $x\in(f,a)$, it will continue to do so. It will never pass between the foci.
Each segment is tangent to a smaller ellipse, confocal with the original
boundary (see figure~\ref{fig:genericorbits}, right panel).
Generically, the orbit is dense in the annular region between the two ellipses.

%%%%%%%%%%%%%%%%%%%%%%%%%%%%%%
%%%%%%%%%%%%%%%%%%%%%%%%%%%%%%

\subsection*{The Homoclinic Orbit: Motion through the Foci.}

\begin{figure}[h]
\begin{center}
\includegraphics[scale=0.667]{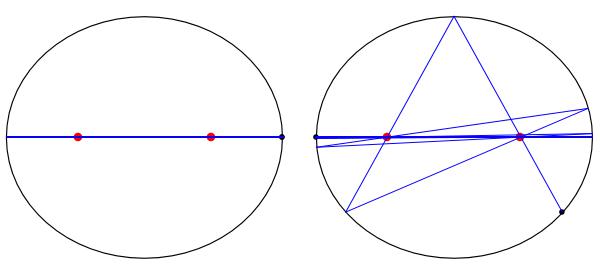}
\caption{Homoclinic orbits. Left: Singular orbit passing through the two foci.
Right: Any orbit through a focus rapidly approaches motion along the major axis.}
\label{fig:focalorbit}
\end{center}
\end{figure}

The two generic families of orbits are separated by the homoclinic orbits,
for which the ball passes through a focus. Due to specular reflection at the boundary,
it will pass through a focus on each subsequent segment of its path.
Moreover, the orbit rapidly approaches horizontal motion back and forth along
the major axis (see figure~\ref{fig:focalorbit}).

%%%%%%%%%%%%%%%%%%%%%%%%%%%%%%
%%%%%%%%%%%%%%%%%%%%%%%%%%%%%%

There is an interesting paradox here: the dynamical behaviour is reversible
in time. However, all orbits through the foci tend to the same limiting
orbit, ultimately bouncing back and forth between the ends $(-a,0)$ and $(+a,0)$
of the ellipse. What if we start on the same trajectory but reverse the
time? We will find that, once again, the orbit will approach horizontal
oscillations. So, starting from motion close to the end state, two things
can happen: either this back-and-forth motion will continue indefinitely
or the $y$-component will gradually increase, reach a maximum and
then die away rapidly. The graph of $y$ against time is reminiscent of
a wave-packet or soliton.

\subsection*{Periodic Orbits.}

For special choices of the initial position and direction of movement,
closed orbits ensue. The simplest are period-2 orbits, where there
are just two points of impact. These are along the major or minor
axes. We have seen that horizontal oscillations are unstable to some
small perturbations: the motion may develop a large component along the
minor axis, and then return to a quasi-horizontal orbit.

There is a simple period-4 orbit touching the ellipse at the four
stationary points, that is the points where the distance from the origin
is maximal or minimal. Now move the initial point $(a,0)$ slightly by varying
$\theta_0=\epsilon$, leaving the direction $\psi_0$ unchanged.
The first impact will move from $(0,b)$ to a nearby point and, after four segments,
the ball will be close to $(a,0)$. It is clear that, by tweaking the
initial angle $\psi_0$, the orbit can be made to return to the initial point
after four segments, thereafter repeating in a period-4 orbit. In fact, this is guaranteed
by Poncelet's closure theorem, discussed below.  It follows from the theorem that there
are period-4 orbits from every starting point.
Indeed, Poncelet's theorem implies that there is a period-$n$ orbit for any
starting point and any $n\ge 3$.

Poncelet's closure theorem, also called Poncelet's porism, is discussed at length in
\cite{Ber05}. A porism is a problem that has either an infinite number of solutions
or none at all. Poncelet's porism states that, given any two conics, if there is a polygon
with vertices on one conic (the outer conic) and all sides tangent to the other one (the
inner conic), then an arbitrary point on the outer conic is a vertex of such a polygon.
For elliptic billiards, this means that if there is a periodic orbit, then
there exist orbits of the same period for any value of the initial position $\theta_0$.
Berger \cite{Ber05} describes Poncelet's result --- which is simply stated
but difficult to prove --- as ``the most beautiful theorem on conics''.

%%%  \subsection*{Central Motion.} %%%  To be completed.

%  \newpage
\subsection*{Phase Portrait.}

The billiard problem may be described as a Hamiltonian dynamical system. Between impacts
with the boundary, the momentum is constant, so the equations of motion are
$$
\mathbf{\dot q} = \mathbf{p} \,,\qquad
\mathbf{\dot p} = \mathbf{0} \,,
$$
representing motion in a straight line at constant speed.
At each impact, the normal component of momentum is reversed while the tangential component
is unchanged.

The dynamics are completely specified by considering the discrete mapping from one bounce
to the next. Let $(x_n,y_n)$ be the $n$-th point of impact, and $(u_n,v_n)$ the velocity
between points $(x_n,y_n)$ and $(x_{n+1},y_{n+1})$. The slope of this segment is
$m_n=v_n/u_n$. Given the values $(x,y;m)_n$ we can calculate the position of the next bounce:
\begin{equation}
x_{n+1} =       -     x_n - \frac{2a^2m_n(y_n-m_nx_n)}{m_n^2a^2+b^2} \,, \qquad
y_{n+1} = \phantom{-} y_n + m_n(x_{n+1}-x_n) \,.
\nonumber
\end{equation}
Then, defining $\nu_{n+1}=(a^2y_{n+1})/(b^2x_{n+1})$, we get
\begin{equation}
m_{n+1} = \frac{2\nu_{n+1}-(1-\nu_{n+1}^2)m_n}{(1-\nu_{n+1}^2)+2\nu_{n+1} m_n} \,.
\nonumber
\end{equation}
We now have $(x,y;m)_{n+1}$.
This discrete map can be iterated to generate the entire orbit.

\begin{figure}[h]
\begin{center}
\includegraphics[scale=0.35]{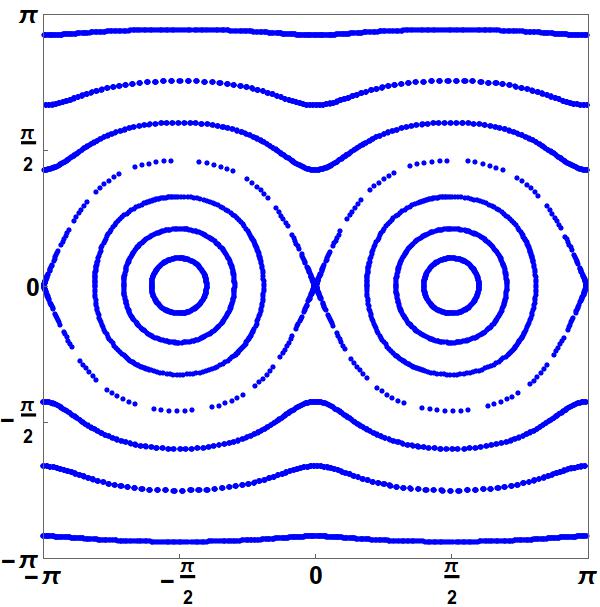}
\caption{Phase plots of ten orbits. Horizontal axis is $\theta\in[-\pi,\pi]$
and vertical axis is $\phi\in[-\pi,\pi]$. The separatrix contains the point $(\theta,\phi)=(0,0)$.}
\label{fig:phaseplots}
\end{center}
\end{figure}

We can determine the motion once the initial position $\theta_0 \in [-\pi,\pi]$ and
\emph{opening angle} $\phi_0 \in [-\pi,\pi]$ are known. Indeed, the continuous system may be represented by
a discrete mapping from one impact to the next:
\begin{equation}
\theta_{n+1} = f(\theta_n,\phi_n) \,, \qquad
\phi_{n+1}   = g(\theta_n,\phi_n) \,.
\nonumber
\end{equation}
Plotting the representative points on a $(\theta,\phi)$ diagram, we obtain a
phase-portrait of the motion. Such portraits are shown in figure~\ref{fig:phaseplots}
for ten choices of initial conditions. Three represent box orbits; these fall within the
separatrix, which includes the point $(\theta,\phi)=(0,0)$.
The remaining six are for clockwise and counter-clockwise loop motions.

There is a striking similarity between these phase plots and the phase portrait for a
simple pendulum, which has librational and rotational motions separated by a homoclinic
orbit that asymptotes to the unstable equilibrium point.

%%%%%%%%%%%%%%%%%%%%%%%%%%%%%%
%%%%%%%%%%%%%%%%%%%%%%%%%%%%%%

%\newpage
\subsection*{Constants of Motion.}

Since the system has no dissipation and since energy is conserved at boundary impacts,
the kinetic energy
$$
T = \half(\dot x^2 + \dot y^2) 
$$
is a constant of the motion. For a circular table it is convenient to use
polar coordinates $(r,\vartheta)$.  Then the Lagrangian may be written
$$
\mathcal{L} = \half(\dot r^2 + r^2 \dot\vartheta^2) - V(r) \,.
$$
Since $\vartheta$ is an ignorable coordinate ($\mathcal{L}$ is independent of $\vartheta$), the
conjugate variable $p_\vartheta=\partial \mathcal{L}/\partial\dot\vartheta=r^2\dot\vartheta$,
the angular momentum about the centre, is conserved.

For an elliptical table, the angular momentum about the centre is no longer conserved.
However, there is another conserved quantity, as we shall now show.  Since the boundary
is an ellipse, it is convenient to use elliptic coordinates $(\xi,\eta)$:
\begin{equation}
x = f \cosh\xi \cos\eta  \,,\qquad y = f \sinh\xi \sin\eta \,.
\label{eq:xieta}
\end{equation}
The components of the velocity $\mathbf{v}=(u,v)$ are then
\begin{eqnarray*}
\dot x &= u &= f \sinh\xi \cos\eta\ \dot\xi - f \cosh\xi \sin\eta\ \dot\eta \\
\dot y &= v &= f \cosh\xi \sin\eta\ \dot\xi + f \sinh\xi \cos\eta\ \dot\eta
\end{eqnarray*}
and the radii from the center and foci are
\begin{eqnarray*}
\mathbf{r_0} = (x  ,y) &=& f (\cosh\xi \cos\eta     , \sinh\xi \sin\eta ) \\
\mathbf{r_1} = (x-f,y) &=& f (\cosh\xi \cos\eta  - 1, \sinh\xi \sin\eta ) \\
\mathbf{r_2} = (x+f,y) &=& f (\cosh\xi \cos\eta  + 1, \sinh\xi \sin\eta )
\end{eqnarray*}
We can now compute the angular momenta $\mathbf{L_1} = \mathbf{r_1\times v}$ 
and $\mathbf{L_2} = \mathbf{r_2\times v}$ about the foci, from which it follows that
\begin{equation}
\mathbf{L_1\cdot L_2} = L_1 L_2 =
f^4(\cosh^2\xi-\cos^2\eta)\bigl[(-\sin^2\eta)\dot\xi^2+(\sinh^2\xi)\dot\eta^2\bigr] \,.
\label{eq:L1dotL2}
\end{equation}
Since there is no cross-term containing $\dot\xi\dot\eta$, the quantity $L_1L_2$ does
not change its value at an impact, where
$(\dot\xi,\dot\eta)\rightarrow(-\dot\xi,\dot\eta)$. Clearly, since segments between
bounces are linear, 
$\mathbf{L_1} = \mathbf{r_1\times v}$ and $\mathbf{L_2} = \mathbf{r_2\times v}$ are
constant along each segment. Thus, $L_1 L_2$ is a constant of the motion \cite{Tab05}

The transformation to elliptic coordinates (\ref{eq:xieta}) can be inverted:
$$
\xi  = \mbox{arccosh}\left(\frac{r_1+r_2}{2f}\right) \,,\qquad
\eta = \arccos\left(\frac{r_2-r_1}{2f}\right) \,.
$$
Constant $\xi$ corresponds to $r_1+r_2$ constant, yielding an ellipse, while
constant $\eta$ corresponds to $r_1-r_2$ constant, yielding a hyperbola.
We note that
$$
r_1 = f(\cosh\xi-\cos\eta) \,,\qquad r_2 = f(\cosh\xi+\cos\eta) \,.
$$
so that 
$$
r_1 r_2 = f^2 (\cosh^2\xi - \cos^2\eta) = f^2 (\sinh^2\xi + \sin^2\eta) \,.
$$
Since a constant product of distances from two points yields a Cassinian oval,
we see that the contours of the function $(\cosh^2\xi - \cos^2\eta)$ are such ovals.
The particular case $r_1 r_2 = f^2$ is the lemniscate of Bernoulli.

It is clear that for loop orbits, $L_1$ and $L_2$ are either both positive or both
negative, so $L_1 L_2$ is positive. For box orbits, which pass between the foci, $L_1$
and $L_2$ are of opposite signs.  For the homoclinic orbit, passing through the foci,
one or other of these components vansihes.
Thus, $L_1 L_2$ acts as a discriminant for the motion:
$$
\mbox{Orbit is}
\begin{cases}
\mbox{Box type}   &\mbox{if\ \ } L_1 L_2 < 0 \\
\mbox{Homoclinic} &\mbox{if\ \ } L_1 L_2 = 0 \\
\mbox{Loop type } &\mbox{if\ \ } L_1 L_2 > 0.
\end{cases}
$$

%%%%%%%%%%%%%%%%%%%%%%%%%%%%%%%%

%  \newpage
\subsection*{Confocal Conics.}

\begin{figure}[h]
\begin{center}
\includegraphics[scale=0.45]{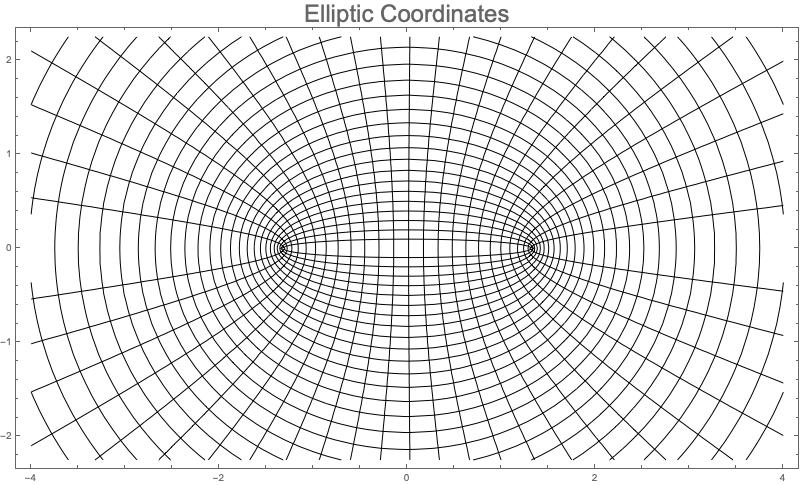}
\caption{Elliptic coordinates $(\xi,\eta)$. The family of ellipses and the family
of hyperbolas each cover the plane. All conics have the same foci.}
\label{fig:ellipcoords}
\end{center}
\end{figure}

%% For problems with boundary conditions on an ellipse,
%%  it is convenient to use elliptic coordinates.
We consider the set of confocal conics
\begin{equation}
\frac{x^2}{a^2+\lambda} + \frac{y^2}{b^2+\lambda} = 1 \,.
\label{eq:confocals}
\end{equation}
The case $\lambda=0$ corresponds to the boundary of the billiard table. The range
$\lambda\in(-b^2,+\infty)$ gives a family of ellipses all having the  same foci, while the range
$\lambda\in(-a^2,-b^2)$ gives a family of confocal hyperbolas orthogonal to the ellipses \cite{Ber05}.
The two families cover the points of the plane twice, and provide an orthogonal coordinate system
(see figure~\ref{fig:ellipcoords}).

Suppose the initial segment lies on the line $y=mx+c$. The condition that this line is tangent to 
a confocal conic (\ref{eq:confocals}) leads to a quadratic equation whose discriminant must vanish,
yielding
\begin{equation}
\lambda = \frac{c^2-(m^2a^2+b^2)}{1+m^2} \,.
\label{eq:lambda}
\end{equation}
But since this is true for all segments of the trajectory, $\lambda$ is an invariant of the motion.

We will show now how this purely geometric result may be interpreted dynamically.
The angular momentum about the centre is $L_0=(xv-yu)$.
The angular momenta about the foci are then $L_1 = L_0-fv$ and $L_2 = L_0+fv$. 
The slope $m$ is related to the components of velocity, $m=v/u$. Using this in
(\ref{eq:lambda}) and recalling that $u^2+v^2=1$, we find that
$$
\lambda = L_0^2 - f^2 v^2 -b^2 =  L_1 L_2 - b^2
$$
Since $L_1 L_2$ is conserved, so is $\lambda$, and every segment of the trajectory is
tangent to the same conic. For $\lambda\in(-b^2,+\infty)$ it is an ellipse, while for
$\lambda\in(-a^2,-b^2)$ it is a hyperbola.

%%%%%%%%%%%%%%%%%%%%%%%%%%%%%%%%%%%%%%%%%%%%%%%%%%%5
\section{Regularizing the Motion: Ballyards}
\label{sec:ballyards}

% \subsection*{Circular Billiard Table.}

The perfectly-reflecting boundaries imply instantaneous changes in momentum at each bounce.
This corresponds to a potential well that is constant in the interior and has a step
discontinuity at the boundary. We can approximate this behaviour by a high-order
polynomial. For a circular table of radius $a$ we take the potential energy to be
$V(r)=V_0(r/a)^N$ where $N$ is a large positive integer. This corresponds to a radially
attractive force 
%%%  $F=-N V_0 (r/a)^{N-1}$
that is negligible in the interior but large 
in a narrow boundary zone. With polar coordinates $(r,\vartheta)$, the conjugate momenta are
$p_r=\dot r$ and $p_\vartheta=r^2\dot\vartheta$ and the Hamiltonian may be written
$$
H = \half(p_r^2+p_\vartheta^2/r^2) + V(r)
$$
Since this is independent of $\vartheta$, the azimuthal momentum $p_\vartheta$ is a constant
of the motion.

%%%%%%%%%%%%%%%%%%%%%%%%%%%%%%
%%%%%%%%%%%%%%%%%%%%%%%%%%%%%%

% \subsection*{Elliptical Billiard Table.}

For an elliptical table, we choose a potential function that rises sharply near the
elliptical boundary.
%%  It is convenient to use elliptic coordinates $(\xi,\eta)$:
%%  \begin{eqnarray*}
%%  x &=& f \cosh\xi \cos\eta \\
%%  y &=& f \sinh\xi \sin\eta
%%  \end{eqnarray*}
%%  
% We assume that the potential $V(\xi)$ depends only on the coordinate $\xi$ and is
% independent of $\eta$.
%
The kinetic energy may be written in elliptic coordinates:
$$
T = \half f^2(\cosh^2\xi-\cos^2\eta)(\dot\xi^2+\dot\eta^2)
$$
The Lagrangian then becomes
\begin{equation}
\mathcal{L} = T - V = 
\half f^2(\cosh^2\xi-\cos^2\eta)(\dot\xi^2+\dot\eta^2) - V(\xi,\eta) \,.
\label{eq:EBlagrangian}
\end{equation}
The conjugate momenta in elliptic coordinates are
$$
p_\xi = \frac{\partial T}{\partial\dot\xi} = f^2(\cosh^2\xi-\cos^2\eta)\dot\xi \,, \qquad
p_\eta = \frac{\partial T}{\partial\dot\eta} = f^2(\cosh^2\xi-\cos^2\eta)\dot\eta \,.
$$
The Hamiltonian may now be written
$$
H =
\displaystyle{\frac{p_\xi^2 + p_\eta^2}{2f^2(\cosh^2\xi-\cos^2\eta)}} + V(\xi,\eta) \,.
$$

\subsection*{A Liouville integrable system.}

\newcommand{\cU}{\mathcal{U}}
\newcommand{\cV}{\mathcal{V}}
\newcommand{\cW}{\mathcal{W}}
\newcommand{\xiB}{\xi_\mathrm{B}}

In 1848 Joseph Liouville identified a broad class of dynamical systems that can be integrated
%  (Whittaker, 1937, \S43). 
\cite[p.~67]{Whi37}.
We confine attention here to systems with two degrees of freedom, with generalised coordinates $(q_1, q_2)$.
If the kinetic and potential energies can be expressed in the form
\begin{equation}
T = \half[\cU_1(q_1)+\cU_2(q_2)]\times[\cV_1(q_1)\dot q_1^2 + \cV_2(q_2)\dot q_2^2] \,,
\qquad
V = \displaystyle{ \frac{\cW_1(q_1)+\cW_2(q_2)}{\cU_1(q_1)+\cU_2(q_2)} } \,.
\label{eq:Louv}
\end{equation}
then the solution can be solved in quadratures. This is proved in Appendix~I,
where explicit expressions for the solution integrals are given.

The kinetic energy term in the Lagrangian (\ref{eq:EBlagrangian}) is of the form
(\ref{eq:Louv}) with
$$
\cU_1(\xi) = f^2\cosh^2\xi \, \qquad \cU_2(\eta) = -f^2\cos^2\eta \, \qquad
\cV_1 \equiv 1 \qquad \cV_2 \equiv 1 \,.
$$
We can express the function $\cU(\xi,\eta)$ in the form
$$
\cU(\xi,\eta) = \cU_1(\xi) + \cU_2(\eta) = r_1 r_2 \,.
$$
For small eccentricity, this depends only weakly on $\eta$.

If the potential energy function is of the form
$V(\xi,\eta) = [\cW_1(\xi)+\cW_2(\eta)]/[\cU_1(\xi)+\cU_2(\eta)]$, the problem
defined by the Lagrangian (\ref{eq:EBlagrangian}) is of Liouville type.
The Hamiltonian may be written as
$$
H =
\displaystyle{
\left[
\frac{\half(p_\xi^2 + p_\eta^2) + (\cW_1(\xi)+\cW_2(\eta))}{f^2(\cosh^2\xi-\cos^2\eta)}
\right] } \,.
$$
It is a constant of the motion, equal in value to the total energy $E$.

\begin{figure}[h]
\begin{center}

\includegraphics[scale=0.15]{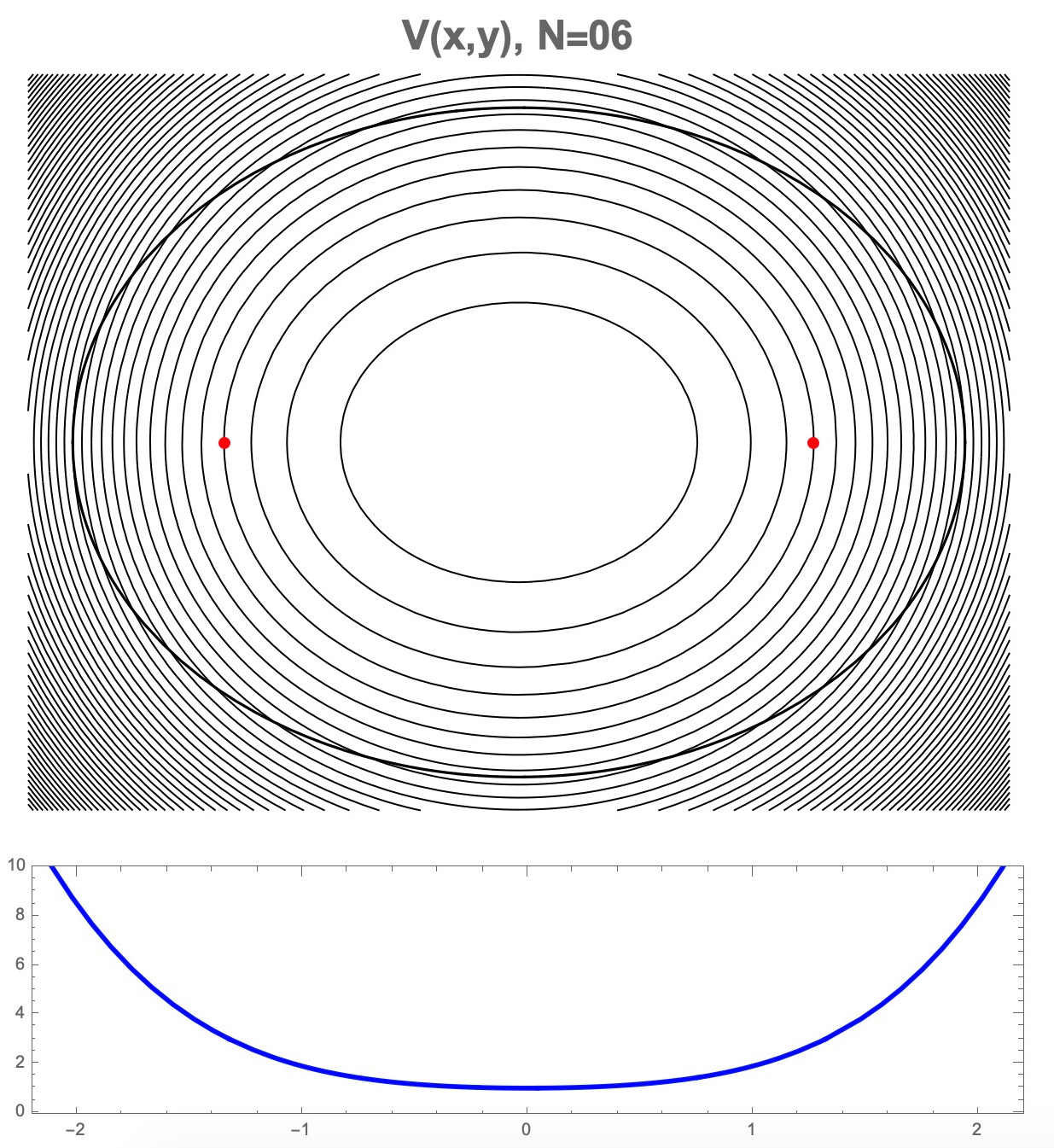}
\includegraphics[scale=0.15]{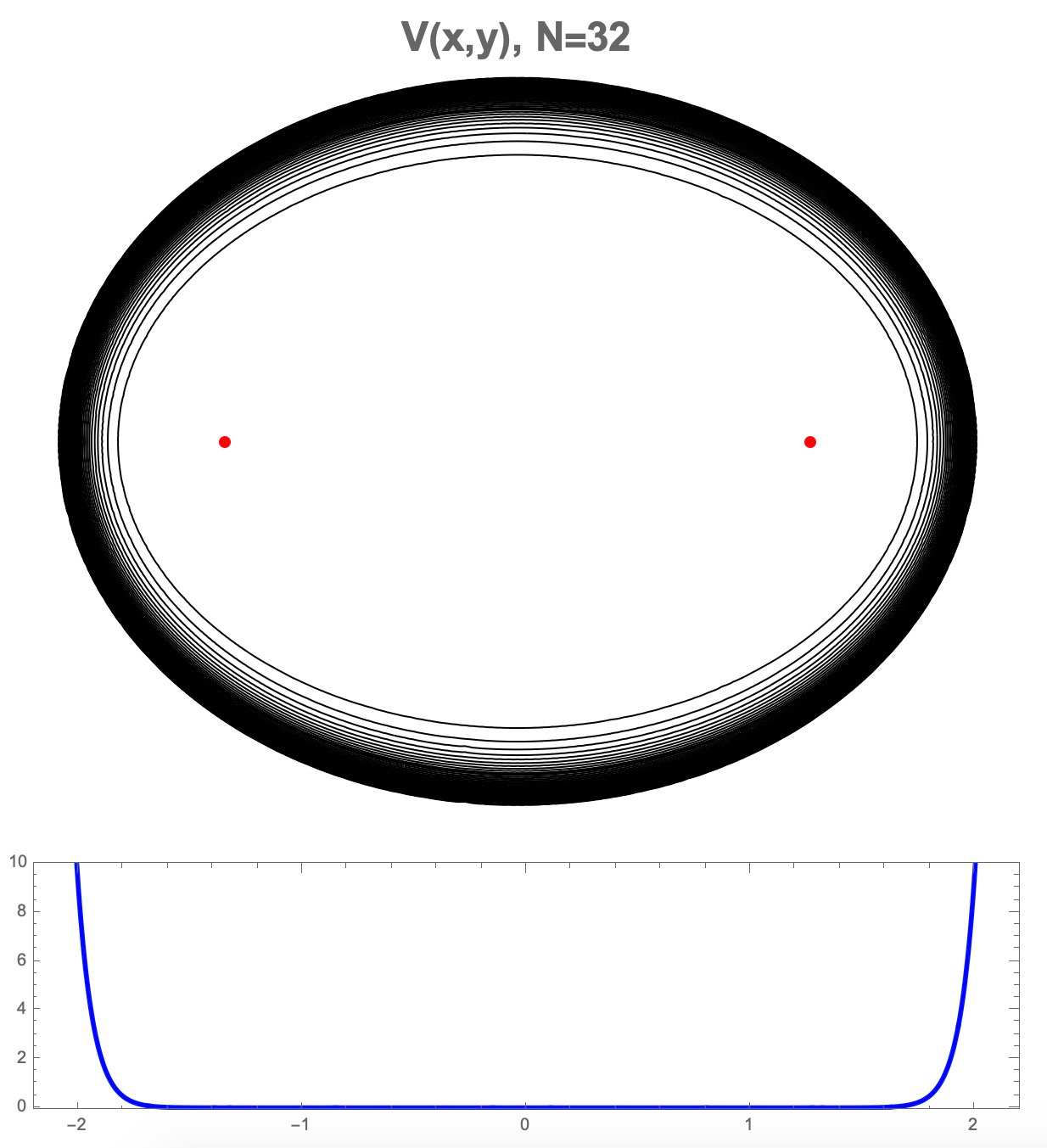}
\caption{Potential energy surface for $N=6$ (left) and $N=32$ (right).
Cross-sections along the major axis are also shown.}
\label{fig:potenergy}
\end{center}
\end{figure}

We seek a potential surface that is close to constant within the elliptical
region defined by $(x/a)^2+(y/b)^2=1$ and that rises rapidly in a boundary zone.
We define the potential surfaces by setting
\begin{equation}
\cW_1(\xi)  =   V_N f^2 \cosh^N\xi \qquad
\cW_2(\eta) = - V_N f^2 \cos^N\eta
\label{eq:cW}
\end{equation}
where $V_N = V_0 \sech^N\xiB $ with $V_0$ a constant, $\xiB$ the value of $\xi$ defining the
reference ellipse ($\cosh\xiB=a/f=1/e$) and $N$ an even integer.
The potential energy function is then
$$
V(\xi,\eta) = 
\frac{\cW_1(\xi)+\cW_2(\eta)}{\cU_1(\xi)+\cU_2(\eta)} =
V_N \left[ \frac{\cosh^{N}\xi - \cos^{N}\eta}{\cosh^2\xi-\cos^2\eta} \right] \,.
$$
Two examples of potential energy surfaces are shown in figure~\ref{fig:potenergy}.

%%%%%%%%%%%%%%%%%%%%%%%%%%%%%%
%%%%%%%%%%%%%%%%%%%%%%%%%%%%%%

For $N=2$ we have
$$
\cW = V_2 f^2 (\cosh^2\xi-\cos^2\eta) \,, \qquad  V \equiv 1 
$$
so that the potential energy is constant. The table is flat and the particle moves at constant
velocity in a straight line; there is no finite boundary. For $N=4$, we have
$$
\cW = V_4 f^2 (\cosh^4\xi-\cos^4\eta) \qquad  V = V_4 (\cosh^2\xi+\cos^2\eta) 
$$
so the potential energy is proportional to $x^2+y^2$. In this case, which corresponds to an
isotropic simple harmonic oscillator in two dimensions, the orbits are closed ellipses.

The system is integrable for any value of $N$ but, for definiteness, we let $N=6$:
$$
\cW = V_6 f^2 (\cosh^6\xi-\cos^6\eta)
 = V_6 f^2 (\cosh^2\xi-\cos^2\eta) (\cosh^4\xi+\cosh^2\xi\cos^2\eta+\cos^4\eta) \,.
$$
The Hamiltonian function for this system is
$$
H =
\displaystyle{
\frac{\half(p_\xi^2 + p_\eta^2)}{f^2(\cosh^2\xi-\cos^2\eta)} 
+ V_6 (\cosh^4\xi+\cosh^2\xi\cos^2\eta+\cos^4\eta)
} \,.
$$

From Equations (\ref{eq:q1eqn}) and (\ref{eq:q2eqn}) in Appendix~I we have
\begin{eqnarray}
\half \cU^2 \dot \xi^2  &=& E \cU_1(\xi)  - \cW_1(\xi)  + \gamma_1 
\label{eq:xieqn} \\
\half \cU^2 \dot \eta^2 &=& E \cU_2(\eta) - \cW_2(\eta) + \gamma_2
\label{eq:etaeqn}
\end{eqnarray}
where $\gamma_1$ and $\gamma_2$ are constants of integration, and $\gamma_1+\gamma_2=0$.

We partition the energy as $E = E_1 + E_2$, where
$$
E_1 = \half\cU(\xi,\eta)\dot \xi^2  + \displaystyle{\frac{\cW_1(\xi)}{\cU(\xi,\eta)}}
\qquad\mbox{and}\qquad
E_2 = \half\cU(\xi,\eta)\dot \eta^2 + \displaystyle{\frac{\cW_2(\eta)}{\cU(\xi,\eta)} \,.}
$$
Then the constants of motion can be written
$$
\gamma_1 = \cU E_1 - E \cU_1 \qquad \gamma_2 = \cU E_2 - E \cU_2  \,.
$$
Note that the components $E_1$ and $E_2$ of energy are not constants.

Equations (\ref{eq:xieqn}) and (\ref{eq:etaeqn}) can be integrated:
\begin{eqnarray}
\int^{\xi} \frac{\cU_1(\xi) \dd\xi}{\sqrt{2[E\cU_1(\xi)-\cW_1(\xi)+\gamma_1}]} &=& \int^t \dd t
\label{eq:solxi}
\\
\int^{\eta} \frac{\cU_2(\eta) \dd \eta}{\sqrt{2[E\cU_2(\eta)-\cW_2(\eta)+\gamma_2}]} &=& \int^t \dd t
\label{eq:soleta}
\end{eqnarray}
Analytical evaluation of these integrals may or may not be possible.
For the case $N=4$, the solutions can be expressed in terms of elliptical integrals.

For the case $N=6$, we get: 
\begin{eqnarray}
\int_{\xi_0}^{\xi} \frac{f^2\cosh^2\xi\,\dd\xi}
                   {\sqrt{2[Ef^2\cosh^2\xi-V_6 f^2\cosh^6\xi+\gamma}]}
                   &=& t-t_0
\label{eq:solxi6}
\\
\int_{\eta_0}^{\eta} \frac{-f^2\cos^2\eta\,\dd \eta}
                     {\sqrt{2[-Ef^2\cos^2(\eta)+V_6 f^2\cos^6\eta-\gamma}]}
                     &=& t-t_0
\label{eq:soleta6}
\end{eqnarray}
where we have written $\gamma=\gamma_1=-\gamma_2$.
These do not have solutions in closed form.

\bigskip
\subsection*{The Angular Momentum Integral.}

For the billiard dynamics, the product of the angular momenta about the foci,
$L_1 L_2$, is an integral of the motion. We seek a corresponding integral for the
general case $N$ of a ballyard. In elliptical coordinates, we can write
$$
L_1 L_2 = f^2 \cU(\xi,\eta)[\sinh^2\xi\ \dot\eta^2 - \sin^2\eta\ \dot\xi^2]
$$
We use (\ref{eq:xieqn}) and (\ref{eq:etaeqn}) to substitute for
$\dot\xi^2$ and $\dot\eta^2$ and, after some manipulation, find that
\begin{equation}
L_1 L_2 + \frac{2f^2(\sinh^2\xi\ \cW_2 - \sin^2\eta\ \cW_1)}{\cU} 
= -2 (f^2 E + \gamma) \,.
\label{eq:LLform}
\end{equation}
Since the right hand side is constant, the same is true of the left hand side.
If we define the quantity
$$
\Lambda(\xi,\eta) =
\frac{2f^2[\sinh^2\xi\ \cW_2(\eta) - \sin^2\eta\ \cW_1(\xi)]}{\cU(\xi,\eta)}
$$
then the relationship (\ref{eq:LLform}) becomes
\begin{equation}
\bbLL \equiv L_1 L_2 + \Lambda = -2 (f^2 E + \gamma) = \mbox{constant} \,,
\label{eq:LLconst}
\end{equation}
and $\bbLL$ is an integral of the motion.

It is easy to show that $\Lambda(\xi,\eta)$ is equal to $\Lambda_0 = -2f^2V_N$,
a constant, on the major axis ($y=0$).
This implies that $L_1 L_2$ is also constant there. But it is clear on physical grounds
that $L_1 L_2 < 0$ on the inter-focal segment $-f < x < f$ and 
$L_1 L_2 > 0$ when $x < -f$ or $x > f$. Therefore, if the trajectory passes through 
the inter-focal segment, it can never cross the major axis outside it. If it crosses
outside, it cannot pass between the foci.
If a trajectory passes through a focus then $L_1 L_2$ must vanish there.
It can never cross the axis at points other than the foci.
This special case of motions $\bbLL = \Lambda_0$ separates the box orbits from the loop
orbits.

We note that as $N\to\infty$, 
$$
\cW_1 = O\left(\frac{\cosh\xi}{\cosh\xiB}\right)^N 
\qquad\mbox{and}\qquad
\cW_2 = O\left(\frac{1}{\cosh\xiB}\right)^N 
$$
so that, for $|\xi|<|\xiB|$, we have $\lim_{N\to\infty}\bbLL = L_1 L_2$.
That is, the integral $\bbLL$ corresponds in this limit to the
product of the angular moments about the foci, the quantity conserved
for motion on a billiard.

Note that (\ref{eq:LLconst}) may be written $\bbLL + 2f^2 E + 2\gamma = 0$, so the three 
constants $E$, $\gamma$ and $\bbLL$ are not independent. Given that the system has two degrees
of freedom, there must be another independent integral of the motion.

\section{Numerical Results}
\label{sec:results}

\begin{figure}[h]
\begin{center}
\includegraphics[scale=0.333]{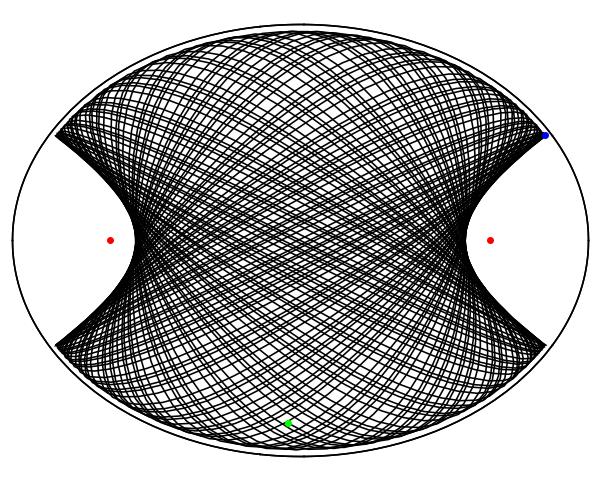}
\includegraphics[scale=0.333]{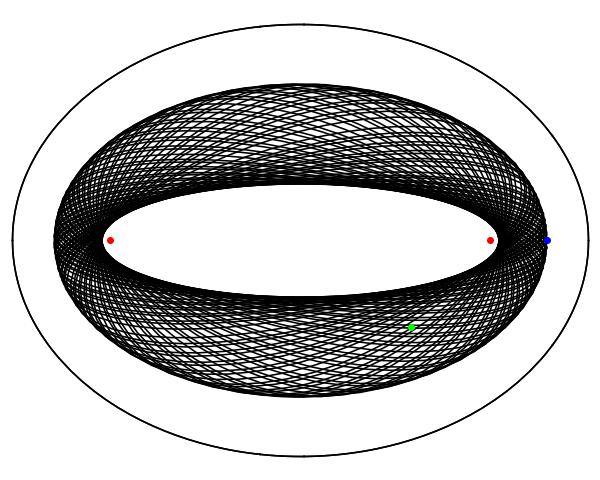}
\caption{Numerical solutions for $N=6$. Left: Box orbit, crossing the major axis between the foci.
Right: Loop orbit, crossing the major axis outside the foci.}
\label{fig:BoxLoopOrbits}
\end{center}
\end{figure}

Numerical integrations confirm the dichotomy between box and loop orbits for the ballyard potentials.
A large number of numerical experiments were performed with $N=6$,
and several for larger values of $N$. Figure~\ref{fig:BoxLoopOrbits} shows
a typical box orbit (left panel) and loop orbit (right panel).
%%%%%%%%%%%%%%%%%%%%%%%%%%%%%%
%%%%%%%%%%%%%%%%%%%%%%%%%%%%%%

\begin{figure}[h]
\begin{center}
\includegraphics[scale=0.333]{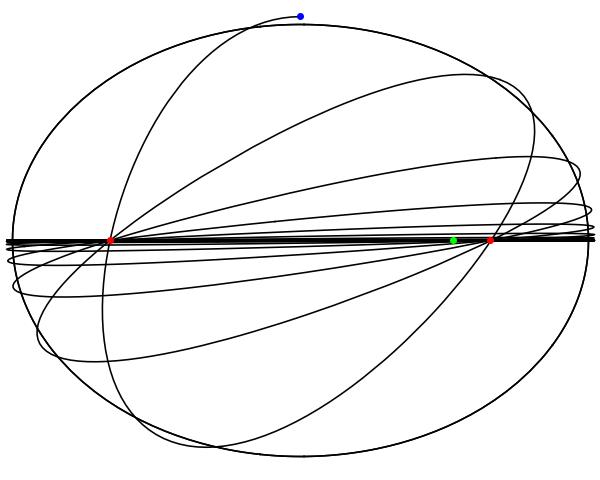}
\caption{Numerical solution ($N=6$) for the homoclinic orbit, crossing the major axis at the foci.}
\label{fig:HeteroOrbit}
\end{center}
\end{figure}

For the special case of orbits passing through the foci, the equations in $(\xi,\eta)$
coordinates are singular. A re-coding using cartesian coordinates enabled numerical
integrations along homoclinic orbits. A typical orbit is shown in figure~\ref{fig:HeteroOrbit}.
The orbit rapidly approaches an oscillation along the major axis. 
%%
%% If the integration is continued, it exhibits intermittent bursts of activity
%% in which it departs for a short time from this motion. These may be presumed
%% to result from instability due to imprecise assignment of the initial conditions
%% for the homoclinic trajectories.

%%%%%%%%%%%%%%%%%%%%%%%%%%%%%%
%%%%%%%%%%%%%%%%%%%%%%%%%%%%%%

\begin{figure}[h]
\begin{center}
\includegraphics[scale=0.333]{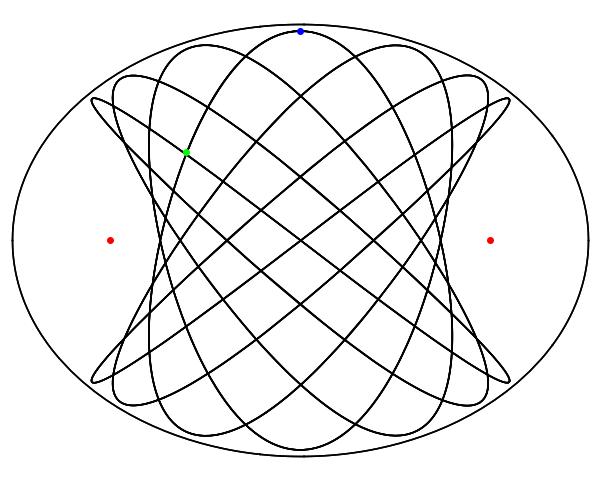}
\includegraphics[scale=0.333]{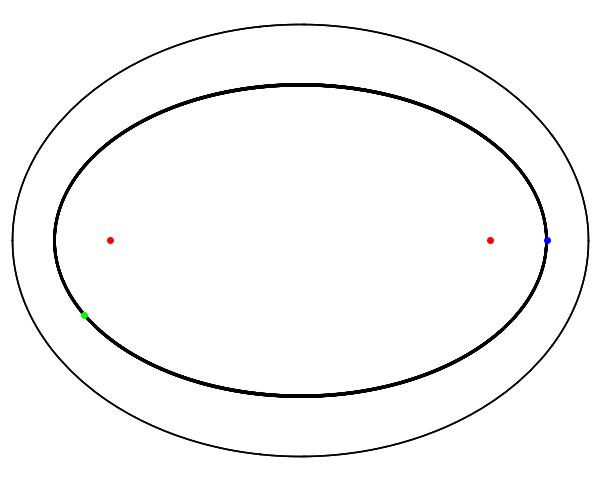}
\caption{Numerical solutions for $N=6$. Left: Periodic box orbit 
Right: Pure elliptic loop orbit.}
\label{fig:closedorbits}
\end{center}
\end{figure}

In general, the orbits are dense in a region bounded by an ellipse and hyperbola (box orbits)
or between two ellipses (loop orbits). However, for delicately chosen initial conditions,
the solutions may be periodic. A periodic box orbit is shown in figure~\ref{fig:closedorbits}
(left panel) and a pure elliptic loop orbit is shown in figure~\ref{fig:closedorbits} (right panel). 
%%%%%%%%%%%%%%%%%%%%%%%%%%%%%%
%%%%%%%%%%%%%%%%%%%%%%%%%%%%%%

\section{Conclusion}
\label{sec:conc}

We have reviewed the well-known problem of motion on an elliptical
billiard, and shown how the dynamics may be regularized by approximating
the flat-bedded, hard-edged billiard by a smooth surface, a ballyard. The
ballyard Lagrangians are of Liouville type and so are completely
integrable. A new constant of the motion ($\bbLL$) was found, allowing us
to show that the trajectories split into boxes and loops, separated by
homoclinic orbits.  The discriminant that determines the character of
the solution is the sign of $L_1 L_2$ on the major axis.

%%%%%%%%%%%%%%%%%%%%%%%%%%%%%%
%%%%%%%%%%%%%%%%%%%%%%%%%%%%%%
\appendix
\section*{Appendix I: Integrable Systems of Liouville Type}
\setcounter{section}{1}

Following
\cite[\S43]{Whi37}, we show how a system of Liouville type, where the
kinetic and potential energies can be expressed in the form
\begin{equation}
T = \half[\cU_1(q_1)+\cU_2(q_2)][\dot q_1^2 + \dot q_2^2] \,, \qquad
V = \displaystyle{\frac{[\cW_1(q_1)+\cW_2(q_2)]}{[\cU_1(q_1)+\cU_2(q_2)]}} \,,
\nonumber
\end{equation}
can be solved in quadratures. Letting $\cU(q_1,q_2)=\cU_1(q_1)+\cU_2(q_2)$ we have
$$
T = \half \cU [\dot q_1^2 + \dot q_2^2] \,, \qquad
V = \displaystyle{ \frac{1}{\cU} [\cW_1(q_1)+\cW_2(q_2)] } \,.
$$
The Lagrangian equation for the coordinate $q_1$ is
$$
\frac{\dd}{\dd t}\left(\frac{\partial T}{\partial\dot q_1}\right)
- \frac{\partial T}{\partial q_1} = - \frac{\partial V}{\partial q_1}
$$
or, more explicitly,
$$
\frac{\dd}{\dd t}\left(\cU \dot q_1 \right)
- \half\displaystyle{\frac{\partial \cU}{\partial q_1}} [\dot q_1^2 + \dot q_2^2] = 
- \displaystyle{ \frac{\partial V}{\partial q_1} }
$$
Multiplying both sides by $2\cU\dot q_1$, this becomes
$$
\frac{\dd}{\dd t}\left(\cU^2 \dot q_1^2 \right)
- \cU\dot q_1 \frac{\partial \cU}{\partial q_1}[\dot q_1^2 + \dot q_2^2] = 
- 2\cU\dot q_1\frac{\partial V}{\partial q_1}
$$
But the expression for total energy $E$ implies
$\cU(\dot q_1^2 + \dot q_2^2) = 2(E - V)$.  Thus the $q_1$-equation may be written
\begin{eqnarray*}
\frac{\dd}{\dd t}\left(\cU^2 \dot q_1^2 \right)
&=& 2(E-V)\dot q_1\frac{\partial \cU}{\partial q_1}
    -2\cU\dot q_1\frac{\partial V}{\partial q_1}    \\
&=& 2\dot q_1 \left[
(E-V)\frac{\partial \cU}{\partial q_1} + \cU\frac{\partial}{\partial q_1}(E-V) \right]  \\
&=& 2\dot q_1 \frac{\partial}{\partial q_1}[(E-V)\cU]  \ \ = \ \  
    2\dot q_1 \frac{\partial}{\partial q_1}[E \cU_1(q_1) - \cW_1(q_1) ]   \\
&=& 2 \frac{\dd}{\dd t} [E \cU_1(q_1) - \cW_1(q_1) ] \,.
\end{eqnarray*}
So we can write
$$
\frac{\dd}{\dd t}\left[\half \cU^2 \dot q_1^2 \right] =
 \frac{\dd}{\dd t} \left[ E \cU_1(q_1) - \cW_1(q_1) \right]  \,.
$$
Integrating this we get
\begin{equation}
\half \cU^2 \dot q_1^2 = E \cU_1(q_1) - \cW_1(q_1) + \gamma_1
\label{eq:q1eqn}
\end{equation}
where $\gamma_1$ is a constant of integration. We obtain a similar equation
\begin{equation}
\half \cU^2 \dot q_2^2 = E \cU_2(q_2) - \cW_2(q_2) + \gamma_2
\label{eq:q2eqn}
\end{equation}
for the $q_2$-component. Adding these together we find that $\gamma_1+\gamma_2=0$.

\subsection*{Solution by Quadratures.}

We partition the energy as $E = E_1 + E_2$, where
$$
E_1 = \half\cU(q_1,q_2)\dot q_1^2 + \displaystyle{\frac{\cW_1(q_1)}{\cU(q_1,q_2)}}
\qquad\mbox{and}\qquad
E_2 = \half\cU(q_1,q_2)\dot q_2^2 + \displaystyle{\frac{\cW_2(q_2)}{\cU(q_1,q_2)} \,.}
$$
Then the constants of motion can be written
\begin{eqnarray*}
\gamma_1 &=& \cU E_1 - E \cU_1 \\
\gamma_2 &=& \cU E_2 - E \cU_2 
\end{eqnarray*}
Note that the components $E_1$ and $E_2$ of energy are not constants.

Equations (\ref{eq:q1eqn}) and (\ref{eq:q2eqn}) can be integrated:
\begin{eqnarray}
\int^{q_1} \frac{\cU_1(q_1) \dd q_1}{\sqrt{2[E\cU_1(q_1)-\cW_1(q_1)+\gamma_1}]} &=& \int^t \dd t
\label{eq:solq1}
\\
\int^{q_2} \frac{\cU_2(q_2) \dd q_2}{\sqrt{2[E\cU_2(q_2)-\cW_2(q_2)+\gamma_2}]} &=& \int^t \dd t
\label{eq:solq2}
\end{eqnarray}
Analytical evaluation of these integrals may or may not be possible.

%%%%%%%%%%%%%%%%%%%%%%%%%%%%%%%%%%%%%%%%%%%%%%%%%%%%%%%%%%%%%%%
% \section*{References}
% \newpage
% \input A900-References.tex

%%%%%%%%%%%%%%%%%%%%%%%%%%%%%%%%%%%%%%%%%%%%%%%%%%%
\end{document}